# Bayesian monitoring of emerging infectious diseases

Pavel Polyakov*, Romulus Breban

Institut Pasteur, Emerging Diseases Epidemiology Unit, Paris, France.

* Corresponding author:

E-mail: pavel.polyakov@pasteur.fr

# Abstract

We define data analyses to monitor a change in $R$, the average number of secondary cases caused by a typical infected individual. The input dataset consists of incident cases partitioned into outbreaks, each initiated from a single index case. We split of the input dataset into two successive subsets, to evaluate two successive $R$ values, according to the Bayesian paradigm. We used the Bayes factor between the model with two different $R$ values and that with a single $R$ value to justify that the change in $R$ is statistically significant. We validated our approach using simulated data, generated using known $R$. In particular, we found that claiming two distinct $R$ values may depend significantly on the number of outbreaks. We then reanalyzed data previously studied by Jansen et al. [Jansen et al. *Science* **301** (5634), 804], concerning the effective reproduction number for measles in the UK, during 1995–2002. Our analyses showed that the 1995–2002 dataset should be divided into two separate subsets for the periods 1995–1998 and 1999–2002. In contrast, Jansen et al. take this splitting point as input of their analysis. Our estimated effective reproduction numbers $R$ are in good agreement with those found by Jansen et al. In conclusion, our methodology for detecting temporal changes in $R$ using outbreak-size data worked satisfactorily with both simulated and real-world data. The methodology may be used for updating $R$ in real time, as surveillance outbreak data become available.

# 1. Introduction

Incidence and prevalence are standard epidemiological indicators, monitored to understand disease dynamic within society [1]. In the case of infectious diseases, it is customary to measure how far an epidemic is from eradication by calculating yet another epidemiological parameter, the basic reproduction number, denoted by $R_0$ (e.g., [2], pp. 4-5). By definition, $R_0$ represents the average number of secondary cases caused by a typical infectious individual in a fully susceptible population. If $R_0$ is larger than 1, then an outbreak becomes an epidemic; otherwise, it goes extinct. If the population is not fully susceptible, then one calculates the effective reproduction number, denoted by $R$ [3, 4]. However, neither $R_0$ nor $R$ is regularly monitored by public health authorities (e.g., [5], pp. 39-65).

$R_0$ depends on a number of factors pertaining to the pathogen-host biology, such as pathogen transmissibility and the natural history of disease. In addition, $R_0$ may depend on factors pertaining to host sociology, such as population density and social awareness about epidemics. All these factors may change with time. Changes in $R_0$ may signal that the pathogen has become more transmissible, virulent or persistent in the population. They may also signal societal re-organization in response to the epidemic dynamic. Particularly important are changes in $R$, which, in addition to changes in $R_0$, may also reflect changes in the susceptibility of the population. Monitoring changes in $R$ may thus be instrumental in determining the success of public health interventions such as mass vaccination [3, 6, 7].

An epidemiological situation that would benefit from $R_0$ and/or $R$ monitoring is that of a zoonotic pathogen repeatedly introduced in a population where it undergoes subcritical transmission. As the pathogen explores more and more hosts, the opportunity for mutations increases, with increasing chance for pandemic strains to occur [8]. Possible applications of this scenario may be the cases of the pre-pandemic severe acute respiratory syndrome [9] and Middle East respiratory syndrome [10].

Another application of $R_0$ and/or $R$ monitoring is to assessing the success of mass vaccination by surveying disease outbreaks before and after a major epidemiological event, such as implementation of mass vaccination [4] and loss of confidence in vaccination programs [6]. Surveillance and contact tracing provides a means of monitoring outbreaks by permanent registration of new cases. For example, monkeypox [11] and non-zoonotic measles [3] are potential candidates for eradication through vaccination. Monkeypox remains worrisome for the possibility that repeated introductions in the human population may yield novel strains of human poxes [12]. Following vaccine licensing in 1963, measles incidence in the United States decreased by more than 95% [13]. Nevertheless, recent $R$ analysis [7] shows the potential for measles to re-emerge and emphasizes the importance of continued surveillance. In this work, we discuss monitoring $R$ using outbreak size surveillance data; only minor modifications are needed to apply our methodology to monitoring $R_0$ of emerging infectious diseases.

There exist two major approaches to estimate $R$ from outbreak size data. In the first approach, $R$ is estimated using the maximum likelihood method. Blumberg et al. [7, 14] used Galton–Watson branching processes to construct the likelihood of observed data consisting of outbreak sizes. By maximizing the likelihood function, they calculated $R$ for monkeypox in the Democratic Republic of Congo (1980–1984) and measles in the United States (2001–2011). Jansen et al. [6] used continuous-time Markov chains to construct the likelihood of observed data. By maximizing the likelihood function, they calculated $R$ for measles outbreaks in the United Kingdom (1995–2002), advocating for an increase in $R$ due to loss of confidence in the vaccination program.

The second approach is based on Bayesian inference, which requires a *prior* distribution describing the current knowledge on the parameter of interest (e.g., $R$) and the likelihood of observing the data set according to a model of choice [15]. As a result, one computes a *posterior* distribution, representing an upgrade of the prior, according to the data. Hence Bayesian inference follows closely the principle of surveillance and learning processes.

Farrington et al. [3] proposed two alternative ways to constructing the posterior probability of $R$ based on (1) outbreak size and (2) outbreak duration. They estimated $R$ for measles (1997–1999) outbreaks in the United States using data from 41 outbreaks caused by a single introduction. Angelov et al. [16] estimated $R$ from a set of clusters sizes, where each cluster consisted from a known number of outbreaks. Following a Bayesian approach, they calculated $R$ for mumps for three different regions of Bulgaria (2005–2008). Yanev et al. [17] described different Bayesian estimators under two different families of loss functions to study multiple outbreaks. They calculated $R$ to describe transmission of smallpox in Europe (1960–1970). Prior knowledge on $R$ was obtained from past data covering the period 1951–1960.

Previous literature focused on extracting a single $R$ value from a given epidemiological dataset. In this work, we describe a methodology to estimate two time-ordered $R$ values from the same set of surveillance data, in the regime of subcritical transmission ($R<1$); however, the generalization to multiple values is straightforward.

## 2. Monitoring temporal changes in $R$

A patient of an outbreak directly infects a number of individuals, so-called secondary cases. In successive generations, each secondary case continues to spread the pathogen, causing secondary cases by themselves. The network of who has infected whom has a tree structure, which, for emerging or re-emerging diseases, may be described using the theory of branching processes [18]. We assume that the data set, denoted by $\mathrm{T}$, is an array of $N$ sizes of trees (i.e., outbreak sizes) ordered by the infection time of the index patient. We analyze the transmission process in terms of effective reproduction number, $R$, which may change with time due to biological (e.g., pathogen transmissibility and virulence) and/or sociological (e.g., frequency of contact between individuals) factors.

## 2.1. General theory

We first briefly present how to compute a single value of $R$ from an epidemiological dataset, in the Bayesian paradigm. To express lack of knowledge about $R$, we use a non-informative improper prior $\pi(R)=1$ $R$ is uniformly distributed from zero to infinity. We construct the likelihood $L(\mathrm{T}|R)$ that the dataset $\mathrm{T}$ is observed, assuming that the effective reproduction number is $R$. Given that the detected trees have sizes $n_1,...,n_N$, ordered according to the date of infection of index patients, we have [14, 18]

$$L(\mathrm{T}|R) = \prod_{i=1}^{N} p(n_i, R), \tag{1}$$

where $p(n_i, R)$ is the probability that the single transmission tree $i$ has size $n_i$ [18]. The posterior distribution $\hat{\pi}(R|\mathrm{T})$ of $R$ for the observed dataset T is calculated according to Bayes' rule [3]

$$\hat{\pi}(R|\mathrm{T}) \propto L(\mathrm{T}|R)\pi(R), \tag{2}$$

which yields

$$\hat{\pi}(R|\mathrm{T}) = L(\mathrm{T}|R) \Big/ \int_0^\infty L(\mathrm{T}|R)dR. \tag{3}$$

The posterior distribution $\hat{\pi}(R|\mathrm{T})$ was used to calculate the average effective reproduction number $\langle R \rangle$ and its 95% credible interval (CI).

The quality of $R$ statistics depends not only on the number of trees but also on their sizes. In short, we summarize the amount of data using the concept of information. According to the definition (e.g., [15], pp. 32), information is given by the logarithm of the probability to observe the given dataset T. In our case, the information, denoted by $I_\mathrm{T}$, is given by the natural logarithm of the likelihood $L(\mathrm{T}|R)$ (c.f. Eq. 1)

$$I_\mathrm{T} = -\ln L(\mathrm{T}|R), \tag{4}$$

so that information is measured in natural units (i.e., nat). Hence, information is presented as the sum of $N$ contributions, one for each tree in T (i.e., an extensive quantity over the number of trees). For a numerical estimate of $I_\mathrm{T}$ corresponding to the dataset, we used the average of $R$ (i.e., $\langle R \rangle$) obtained from the Bayesian framework.

Monitoring the pandemic risk of emerging infectious diseases, one extracts at least two time-ordered $R$ values from the same dataset $T$. Hence, the dataset $T$ is divided into two time-ordered subsets $T_a$ and $T_b = T \setminus T_a$. The Bayesian approach described above (c.f. Eq. 3) is then applied independently to each subset $T_{a,b}$ to estimate two effective reproduction numbers $R^{a,b}$. Obviously, the estimates $R^{a,b}$ depend on the selection of $T_{a,b}$. For example, if $T_a$ is nearly all $T$ then $R^b$ is badly estimated.

To justify the choice of extracting two $R$ values from $T$, we proceed as follows. We denote by $H_T$ the model with a single $R$ estimate and by $H_{a,b}$ the model with two $R$ estimates. Given $R^{a,b}$, the likelihood for the dataset $T_a \cup T_b$ is given by $L(T_a|R^a)L(T_b|R^b)$ [c.f., Eq. 1 for $L(T|R)$]. We evaluate whether $H_{a,b}$ is more plausible than $H_T$ by calculating the Bayes factor [19], using the corresponding likelihoods. When our initial beliefs are a priori equally probable, $\mathrm{pr}(H_{a,b}) = \mathrm{pr}(H_T)$, the Bayes factor

$$B(T_a) = \frac{L(T_a|R^a)L(T_b|R^b)}{L(T|R)} \qquad (5)$$

expresses how well the observed data were predicted by $H_{a,b}$, compared to $H_T$; i.e., the higher the value of $B(T_a)$, the more is justified to extract two $R$ values rather than one from $T$. Kaas et al. [19] provided an interpretation of the strength of the second model $H_{a,b}$ in terms of four categories according to the gradation of $2\ln B(T_a)$. They suggested a very strong preference for $H_{a,b}$ if $2\ln B(T_a) > 10$. In our model, we make the same decision. We extract two $R^{a,b}$ values when $2\ln B(T_a) > 10$; otherwise, we extract a single value of $R$ from $T$.

Our $R$ analysis is performed according to the following steps. For every $i = 1,...,(N-1)$:

(1) Let $T_a$ consist of the first $i$ trees in $T$ and $T_b = T \setminus T_a$;

(2) Estimate the pair of posterior distributions $\hat{\pi}(R^{a,b} | T_{a,b})$ using Eq. 3; Numerical integration of the normalization constant was performed using the trapeze rule.

(3) Use the average $\langle R \rangle^{a,b}$ as an estimate of the parameters $R^{a,b}$.

(4) Calculate $B(T_a)$ using Eq. 5.

Finally, we denote by $T^*_{a,b}$ the sets $T_{a,b}$ which yield the largest value of $2\ln B(T_a)$. If $2\ln B(T^*_a) > 10$, we accept $\langle R \rangle^{a,b}$ as the best estimates of effective reproduction numbers on $T^*_{a,b}$ and we denote them by $\langle R \rangle^{a,b*}$; otherwise, we calculate a single $R$ value on $T$ using Eq. 3.

## 2.2. Numerical tests using synthetic data

Synthetic data consisting of arrays of sizes of transmission trees were simulated, assuming that the number of cases caused by each infected individual is a Poisson deviate with average $R$. Tree sizes are random integers given by a distribution $p(n_i, R)$, obtained from the probability-generation function for the Galton–Watson branching processes according to Pintman [18]. For the Poisson offspring distribution, we obtain

$$p(n_i, R) = \frac{(n_i R)^{n_i - 1} \exp(-n_i R)}{n_i!}. \qquad (6)$$

The synthetic dataset, consisting of sizes of $N$ transmission trees with known $R$, was used to validate our methodology of estimating $R^{a,b}$. The analytical approach according to Eq. 3 yields

$$\langle R \rangle = 1 - 1/\bar{n} + 1/(\bar{n}N), \qquad (7)$$

where $\bar{n}$ is the average size of the trees in $T$. The corresponding standard deviation

$$std(R) = \sqrt{\frac{\langle R \rangle}{\bar{n}N}}, \qquad (8)$$

becomes negligible when $N \gg 1$. Eq. 7 gives $R=1$ if T consists of a single tree. For large tree number in T, the third term goes to zero and Eq. 7 becomes identical to the formula derived from the maximum likelihood $L(T|R)$ method.

Using Eq. 7, we estimate the change of $R$ in real time, as new epidemiological data become available. Suppose we add a new tree of size $n$ to T. The corresponding change in $R$,

$$\delta R = \frac{n(1-1/N)-\bar{n}}{\bar{n}(\bar{n}N+n)}, \tag{9}$$

shows that, for a sufficiently large tree number in the dataset $N \gg 1$, the sign of the change in $R$ is determined by the difference between the size $n$ of the newly added tree and $\bar{n}$.

We now proceed with the discussion of our simulations. In the first example, we generated a homogeneous dataset T, consisting of 100 trees with $R$ equal to 0.6. Figure 1(a) shows $2\ln B(T_a)$ as a function of information (c.f. Eq. 4) in the first subset $T_a$. The observed maximum is ~3, indicating weak justification for calculating two $R$ values. Hence, we concluded that this dataset should be assigned a single $R$ estimate.

In the second example, we assumed a stepwise increase of $R$, modeling adaptation of the pathogen to human-to-human transmission. We generated a non-homogeneous synthetic dataset where 50 trees were generated with $R^a = 0.6$, while the remaining 50 trees were generated with $R^b = 0.85$. The parameter $2\ln B(T_a)$ (c.f. Fig. 1(b)) reaches its maximum at the 55$^{th}$ tree. Contrary to the previous example, the observed maximum of $2\ln B(T_a)$ is ~14, suggesting strong justification for calculating $R^{a,b}$. The best $R^{a,b}$ estimates, denoted by $R^{a,b*}$, have non-overlapping error bars (95% CI) and are in satisfactory agreement with the numerical choices for the parameters when $2\ln B(T_a) > 10$ (c.f. Fig. 1(c)).

In order to clarify the minimal size of the initial dataset required to get reliable estimations of $R^{a,b}$, we performed evaluations of $R^{a,b*}$ for statistically independent synthetic datasets $T$, containing from 2 to 200 trees. For each size of $T$, we averaged 120 independent realizations to alleviate stochastic effects. Figure 2(a) shows the average $2\ln B(T_a^*)$ and its 95% confidence interval as a function of the total number of trees in the homogeneous datasets, when $R$ equal to 0.6. The values are bellow 10 and the featured dependence is not particularly sensitive to the number of trees in $T$.

The situation is quite different for non-homogeneous datasets. As an example, we generated a set of trees $T$ with $R$ equal to 0.6 and 0.85 for the first and second halves, respectively. The results are presented in Fig. 2(b), where the average $2\ln B(T_a^*)$ increases with the number of trees in $T$. For a low number of trees (up to ~100 trees in $T$), the average of $2\ln B(T_a^*)$ is less than 10, indicating no preference to estimate two $R$ values. For high number of trees, the average of $2\ln B(\mathrm{T}_a^*)$ is larger than 10, demonstrating that the model with two $R$ values is superior. For small sizes of $\mathrm{T}$ (c.f. Fig. 2(c)), average estimates of $R^{a,b}$ display strong fluctuations. Starting with ~50 trees per dataset $\mathrm{T}$, $R$ estimates are neater. With further increasing the size of $\mathrm{T}$, $R$ estimates are more grouped around the exact values and the error bars (95% CI) are smaller.

## 2.3. Application to epidemiological data

We applied our method of evaluating two $R$ values from an epidemiological dataset. We aimed to reproduce recent results obtained by Jansen et al. [6], concerning the transmission dynamics of measles in the UK during 1995–2002. Although measles-elimination programs were set worldwide, measles eradication has not yet been achieved. In the late nineties, the safety of a combined measles-mumps-rubella (MMR) vaccine became controversial, which resulted in decreased uptake of the MMR vaccine subsequent to 1998. As a consequence, measles outbreaks increased in sizes.

Jansen et al. [6] calculated two effective reproduction numbers $R$ for the UK, regarding the periods 1995–1998 and 1999–2002, respectively. They found that $R$ increased significantly, from 0.47 to 0.82. The epidemiological data consisted of measles cases grouped into outbreaks of size 2 or more. We accounted for the left censoring of the outbreak-size data by renormalizing the probability of observing outbreaks $p(n_i,R)$ as $p(n_i,R)/(1-p(1,R))$, where Eq. 6 provided the probability model. The results are presented in Fig. 3. The parameter $2\ln B(T_a)$ shows a marked maximum (c.f., Fig. 3(a)) at the 35$^{th}$ outbreak, the latest one of 1998. The magnitude of $2\ln B(T_a)$ is ~14, justifying the split of the 1995–2002 dataset into two separate subsets for the periods 1995–1998 and 1999–2002. Jansen et al. [6] used the same split of the dataset, based on information about measles-vaccination coverage and the MMR controversy. Our analysis yielded the same conclusion, based on the measles outbreak data alone.

Figure 3(b) shows $R^{a,b}$ as a function of $2\ln B(T_a)$. Both $R^{a,b}$ have nearly constant values for $2\ln B(T_a) > 10$. For the maximum value of $2\ln B(T_a) > 10$, we evaluated $R^{a,b*}$ as 0.54 (90%CI 0.43–0.66) and 0.86 (90%CI 0.79–0.93), respectively. The quantitative difference with the results by Jansen et al. [6] ($R^a = 0.47$ (90%CI 0.36–0.55) and $R^b = 0.82$ (90%CI 0.71–0.87)) is not significant ($p_a = 0.92$ and $p_b = 0.96$ for $R^{a,b}$, respectively) and may be explained by the difference in the methods of data analysis.

## 3. Discussion

We propose a method to monitor the effective reproduction number of infectious diseases with sub-threshold transmission. The method may apply to alert and surveillance systems of diseases emerging and/or re-emerging from natural reservoirs. In this case, monitoring an increase in $R_0$ and/or $R$ may be used to determine the implementation of public health intervention. The method may also be used to asses the effectiveness of public-health programs designed for disease elimination. In this case, evaluating $R$ before and after implementing intervention may confirm the performance of the public-health program.

We validated our method using synthetic data, of which we presented a few simulations. Lastly, we reanalyzed data previously studied by Jansen et al. [6], concerning the transmission of measles in the UK during 1995–2002. While our $R$ findings are similar, we also extracted from the epidemiological data, the time when rumors on the MMR vaccine started to impact on measles vaccination. Of note, in a previous publication, Blumberg et al analyzed the transmissibility of measles in the US (1997-1999) and Canada (1998-2001), estimating two values of R from two distinct datasets. However, further epidemiological assumptions are required for a direct comparison of R resulting from the two analyses.

Our model has several limitations. Epidemiological data are often meant for estimating incidence or prevalence (e.g., [1], [5] pp. 39-65). However, our study requires a particular type of longitudinal data, consisting of outbreak sizes where the outbreaks are ordered according to the date of infection of index patients. These data may result from close surveillance of emerging or re-emerging infections. However, outbreaks may be clustered in time and space and difficult to tell apart. If the number of index patients in each cluster is determined, the model could be amended by using the Borel-Tanner distribution [20] for $p(n_i, R)$ in Eq. 6.

Our model provides monitoring of $R$ in the regime of subcritical transmission (i.e., $R<1$); other methods may be used for estimating $R$ in the regime of supercritical transmission [21]. The quality of $R$ estimation in our model depends on the number of outbreaks; see Fig. 3c. Therefore, the number of changes in $R$ that can be estimated from a dataset depends on the number of outbreaks. We addressed the case of detecting a single change in $R$ by solving a one-dimensional maximization problem for the Bayes factor $B(T_a)$ (c.f., Eq. 5). There is no coincidence that (global) maximization problems are divided between one- and multi-dimensional, the first class of problems being significantly easier. However, a number of numerical algorithms are readily available to address maximization in several dimensions [22]. In our case, such algorithms would have to face additional difficulties, inherent to the stochastic nature of the data modeling.

In conclusion, our work proposes a novel method to monitor changes in the effective reproductive number from an epidemiological dataset consisting solely of outbreak sizes.

# Acknowledgments

We thank Prof V.A.A. Jansen (Royal Holloway University of London) for the epidemiological data. P.P. is grateful for financing in the form of a postdoctoral scholarship from Agence Nationale de la Recherche (Labex Integrative Biology of Emerging Infectious Diseases).

# Figures

**Fig. 1.** $R^{a,b}$ **values for synthetic datasets** $T$. We considered simulated datasets consisting of 100 outbreak sizes; the vertical arrows show the position of the 50th tree. Panel (a) corresponds to $R=0.6$ and shows $2\ln B(T_a)$ as a function of information in $T_a$, as $T_a$ increases from including the 1st tree only to including the first $N-1$ trees. The maximum of $2\ln B(T_a)$ is bellow 10, suggesting preference for a single $R$ estimate over the whole dataset. Panel (b) is similar to panel (a), except T consisted of 50 trees with $R^a=0.6$ and 50 trees with $R^b=0.85$. Values of $2\ln B(T_a)$ larger than 10 show where the dataset T can be split into $T_{a,b}$ with justification for evaluating $R^{a,b}$. The corresponding $R^{a,b}$ values are shown in the panel (c) as a function of $2\ln B(T_a)$.

**Fig. 2. The impact of amount of data on** $R$ **evaluation.** We considered 120 independent realizations of synthetic datasets with number of outbreaks between 2 and 200. Panels (a) and (b) show the average of $2\ln B(T_a^*)$ and its 95% confidence interval as a function of tree number for homogeneous and non-homogeneous $T$, respectively. In particular, the panels are as follows. (a) All trees in T are generated with $R=0.6$. The average $2\ln B(T_a^*)<10$ recommends one $R$ evaluation from $T$. (b) The firsts half of the number of outbreaks in $T$ has $R^a=0.6$, while the second half has $R^b=0.85$. The average $2\ln B(T_a^*)$ increases

with the number of trees in $T$. Once above 10 (i.e., starting from datasets of 100 outbreaks or more), it is justified estimating $R^{a,b}$. The corresponding $\langle R \rangle^{a,b\,*}$ values, calculated for each dataset $T$, are shown in the panel (c), as a function of the number of outbreaks in $T$.

**Fig. 3. $R^{a,b}$ values for measles in the UK during 1995–2002.** The dataset consists of 78 trees, Jansen et al. [6]. (a) $2\ln B(T_a)$ versus information in $T_a$. Since $2\ln B(T_a) > 10$, we evaluate $R^{a,b}$. The vertical arrow indicates the best splitting point of $T$, determined by the maximum of $2\ln B(T_a)$. (b) $\langle R \rangle^{a,b\,*}$ values, plotted as a function of $2\ln B(T_a)$, were calculated for all frames $T_{a,b}$ where $2\ln B(T_a) > 10$.

**Figures**

Figure 1

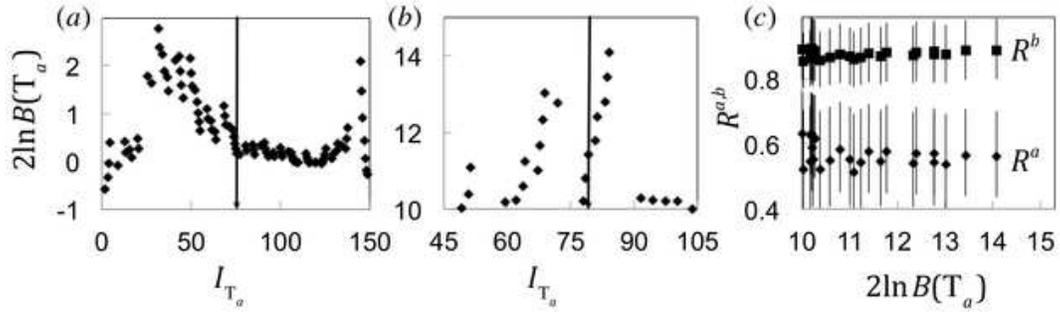

Figure 2

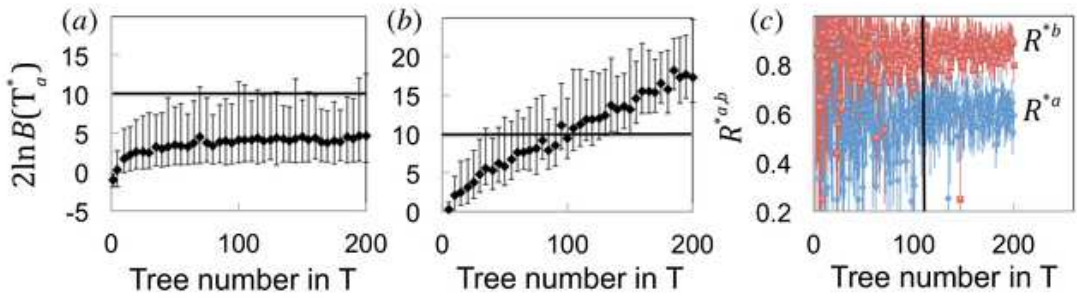

Figure 3

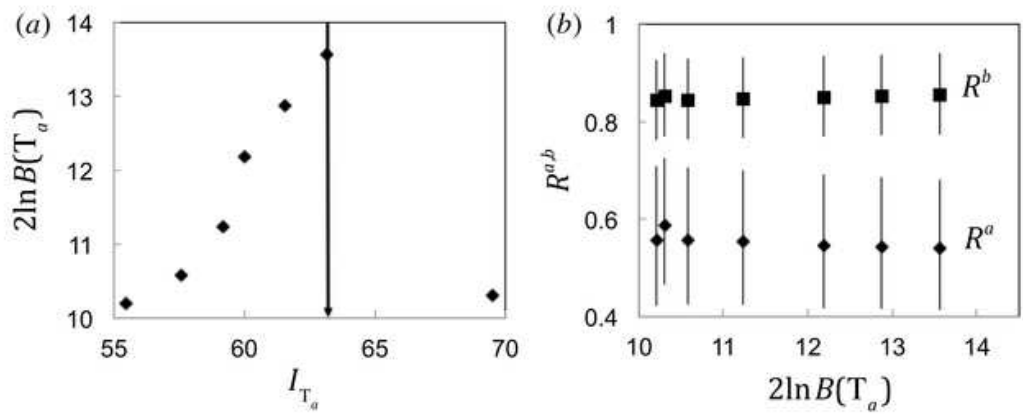